\documentclass[11pt,english,svgnames]{article}
\usepackage[english]{babel}
\usepackage[latin1]{inputenc}
\usepackage[a4paper, body={16cm,23.7cm}]{geometry}
\usepackage{booktabs}
\usepackage{amssymb,amsthm,amsfonts,amsmath}
\usepackage{enumerate}
\usepackage{float,graphicx,xcolor,tikz,caption,subcaption}

\usetikzlibrary{positioning}

\newtheorem{theorem}{Theorem}

\newtheorem{proposition}{Proposition}

\begin{document}

\title{\textbf{Redistribution with Needs}\thanks{We thank an associate editor and two anonymous referees for helpful comments and suggestions. We also acknowledge contributions from participants at the 2022 Economic Design Meeting (Padova), the 2022 Economic Theory meeting (Canberra), the 2022 Workshop on Mechanism Design and Welfare Economics (Malaga), the 6th Spain-Japan Meeting on Economic Theory and the 2023 Workshop on Collective Decisions (Granada), as well as seminar audiences at UNSW, Universidad de Vigo, Universidad del Pais Vasco and Waseda University. Financial support from the Spanish Ministry of Economics and Competitiveness, through the research projects PID2020-115011GB-I00, PID2020-114309GB-I00 and Junta de Andaluc\'{\i}a through grants P18-FR-2933 and A-SEJ-14-UGR20 is gratefully acknowledged.}
}
\author{Ricardo Mart\'{\i}nez\thanks{ Universidad de Granada,
Spain. email: ricardomartinez@ugr.es} \and Juan D. Moreno-Ternero\thanks{Universidad Pablo de Olavide, Spain. email: jdmoreno@upo.es}}
\maketitle

\begin{abstract}
We take an axiomatic approach to study redistribution problems when agents report income and needs. We formalize axioms reflecting ethical and operational principles such as additivity, impartiality and individual rationality. Different combinations of those axioms characterize three focal rules (\textit{laissez faire}, \textit{full redistribution}, and \textit{need-adjusted full redistribution}) as well as compromises among them. 
We also uncover the structure of those compromises exploring the Lorenz dominance criterion as well as majority voting.
Our analysis provides an axiomatic justification for a linear income tax system. 
We conclude our analysis resorting to Eurostat's Household Budget Survey from where we illustrate the different redistribution patterns accounting for needs across European countries. 
\end{abstract}

\bigskip

\medskip {}


\newpage

\section{Introduction}

The redistributive effect of income taxation has long been established. In this paper, we focus on the specific problem of redistribution when agents report income and needs. We approach this problem normatively, endorsing the axiomatic method. To do so, we consider a stylized model in which an income profile reflects the taxable and observable income of a group of agents, and a needs profile reflects the amounts those agents objectively need. Our model is flexible enough to accommodate agents with negative (pre-tax) incomes or with needs above their (positive) incomes. A plausible interpretation for needs is \textit{basic consumption} (e.g., Stone, 1954), which agents need to survive. 
This justifies to consider need as independent of income. The issue is to construct rules that transform an income profile into another, without wasting in the redistribution process. 
The model is, therefore, a generalization of the income redistribution model introduced by Ju et al., (2007) in which the problem is addressed without considering needs.\footnote{See also Alm\r{a}s et al., (2011), Casajus (2015), Chambers and Moreno-Ternero (2021), or Martinez and Moreno-Ternero (2022) for recent contributions on that model.} 
It is also reminiscent of the seminal model introduced by O'Neill (1982) to analyze claims problems, surveyed by Thomson (2019).\footnote{More precisely, when claims problems are amended via the so-called \textit{baselines} (e.g., Hougaard at al., 2012) that could also be considered as needs. Recently, Billette de Villemeur and Leroux (2023) have also extended the standard cost-sharing problems to account for needs.} Young (1988, 1990) reinterpreted that same model to analyze taxation problems, focussing on the principle of equal sacrifice.\footnote{See also Ju and Moreno-Ternero (2011, 2023).} 
In our model, the tax revenue to be raised is zero. And income profiles are amended with needs, which might play a role in the redistribution process.

We start our axiomatic analysis considering three basic and intuitive axioms for this model: \textit{equal treatment of equals}, \textit{continuity} and \textit{additivity}. The first one is a standard notion of impartiality, a fundamental requirement in the theory of justice (e.g., Moreno-Ternero and Roemer, 2006). It states that if two agents are equal (they have equal income and equal needs), then they receive the same amounts. The other two axioms are standard notions in axiomatic work that exclude possible arbitrariness in the allocation process, while conveying a form of simplicity. They state, respectively, that small changes in the data can only produce small changes in the solution, and that the process is additive with respect to incomes and needs.\footnote{\textit{Additivity} can be traced back to Shapley (1953). De Clippel and Rozen (2022) provide strong evidence supporting this axiom. We discuss its normative underpinnings in our last section.} Our first result (Theorem 1) characterizes all the rules satisfying the three axioms. 
They are rules that emerge as certain linear combinations of three fundamental rules: \textit{Laissez Faire}; \textit{Full Redistribution} and \textit{Need-Adjusted Full Redistribution}. The first two rules are self-explanatory. The third rule is the one that assigns needs first, and then redistributes equally the residual (overall) income.

We then consider several additional natural axioms. On the one hand, we present axioms reflecting natural lower bounds, which have a long tradition of use in the literature on fair allocation (e.g., Thomson, 2011).\footnote{This is a long-standing concern in political philosophy that can be traced back to Rawls (1971) and, more recently, to van Parijs and Vanderborght (2017) or Chambers and Moreno-Ternero (2017).} On the other hand, we present axioms reflecting the principles of monotonicity and individual rationality. We show that combining some of these axioms with some of the three mentioned above allows us to characterize natural (pairwise) compromises (in the form of convex combinations) of the three fundamental rules (Theorem 2), as well as the three rules themselves independently (Theorem 3).

We complement our axiomatic analysis in three ways.

First, we take a decentralized approach to single out a specific rule within the families we characterize. Following the long tradition of voting for income tax schedules (e.g., Romer, 1975; Roberts, 1977; Calabrese, 2007) we study whether the choice of a rule within a family could be made by means of simple majority voting. Due to the overwhelming existence of majority cycles (e.g., Greenberg, 1979), one should normally not expect a positive answer to this question. Somewhat surprisingly, we do obtain the majority voting equilibria for each of the three families providing pairwise compromises among the fundamental rules (Propositions 1-3). 

Second, we explore the structure of the characterized families resorting to \textit{Lorenz domination}, the most fundamental concept of income inequality (e.g., Dasgupta et al., 1973). We say that a rule \textit{Lorenz dominates} another rule if, for each problem, the allocation the former yields is greater than the allocation the latter yields in the Lorenz ordering. As the Lorenz criterion is a partial ordering, one might not expect to be able to perform many comparisons of rules. Nevertheless, we do obtain that two of our families of rules can be fully ranked according to the Lorenz criterion (Proposition 4).

Third, we connect our work with the well-known literature on optimal income taxation pioneered by Mirrlees (1971).\footnote{We thank an anonymous referee for suggesting this connection.} 
We can naturally define the taxation rule associated to a redistribution rule in our setting. 
Consequently, we characterize the taxation rules that satisfy \textit{equal treatment of equals}, \textit{continuity}, and \textit{additivity} (Theorem \ref{tax}). They are linear tax systems that decompose into an income tax rate, a tax deduction, and a lump-sum subsidy for need. 
Resorting to the results on Lorenz domination mentioned above, we deduce the following: as the marginal income tax rate increases, the degree of inequality in the family of taxation rules decreases. 
This confirms the traditional view in public economics stating that, with income taxation limited to a linear system, the optimal (marginal) income tax rate is greater as social concern for inequality increases (e.g., Atkinson and Stiglitz, 1980).

We conclude our paper resorting to Eurostat's Household Budget Survey to provide an illustration of our analysis. This survey yields a picture of living conditions across the European Union, gathering information on households' expenditure on goods and services at the national level. Based on it, we obtain the distributions of income and needs of European households and represent the patterns that different nations exhibit, with respect to the redistribution rules we characterize in our analysis.

The rest of the paper is organized as follows. Section 2 presents the model and basic definitions. 
Section 3 is devoted to our axiomatic analysis, providing our characterization results. Section 4 explores the decentralized approach in which agents vote for rules. Section 5 uncovers the structure of the families of rules we characterize via the Lorenz criterion. Section 6 explores the connection with the literature on taxation. Section 7 presents our empirical illustration. Section 8 concludes. For a smooth passage all proofs are relegated to an Appendix.


\section{The model}
Let $N=\{1,\dots, n\}$ be a set of individuals, or \textbf{agents}. For each $i\in N$, let $y_{i}\in \mathbb{R}$ and $z_{i}\in \mathbb{R}_+$ be $i$'s \textbf{income} and \textbf{need}, respectively.\footnote{Note that incomes may be negative.} We denote by $y\equiv (y_i)_{i\in N}$ and $z\equiv (z_i)_{i\in N}$ the corresponding profiles of incomes and needs. The aggregate income is $Y \equiv \sum_{i \in N} y_{i}$, while the aggregate need is $Z \equiv \sum_{i \in N} z_{i}$. A redistribution problem with needs (in short, a \textbf{problem}) is a pair $(y,z) \in \mathbb{R}^N \times \mathbb{R}_+^N$. Let $\mathcal{D}^{N}$ be the domain of all problems. We shall also consider a variable-population generalization of the model. Then, there is a set of potential agents, which are indexed by the natural numbers $\mathbb{N}$.  Let $\mathcal{N}$ be the set of finite subsets of $\mathbb{N}$, with generic element $N$. Let $\mathcal{D} \equiv \bigcup _{N\in \mathcal{N}}\mathcal{D}^{N}$ be the domain of all problems.

Given a problem $(y,z) \in \mathcal{D}^{N}$, an \textbf{allocation} is a vector of real numbers $x \equiv (x_i)_{i \in N} \in \mathbb{R}^N$ such that $\sum_{i \in N} x_i = Y$. Let $X(y,z)$ denote the set of all allocations for the problem $(y,z)$. An allocation rule, or simply a \textbf{rule}, is a mapping $R:\mathcal{D} \longrightarrow \bigcup_{N \in \mathcal{N}} \mathbb{R}^N$ that selects, for each problem $(y,z) \in \mathcal{D}^{N}$, a unique allocation $R(y,z) \in X(y,z)$.

We consider three basic axioms for rules. 

First, we consider a standard notion of \textit{impartiality}. 
It states that if two agents are equal (they have equal income and equal needs), then they receive the same amounts in the allocation.

\textbf{Equal treatment of equals}. For each $(y,z) \in \mathcal{D}^{N}$ and each pair $i,j \in N$ such that $y_i=y_j$ and $z_i=z_j$, $R_i(y,z)=R_j(y,z)$.

The second axiom conveys the standard requirement that small changes in the data of the problem should not lead to large changes in the allocation.

\textbf{Continuity}. For each sequence $\{(y^\nu,z^\nu)\}$ of problems in $\mathcal{D}^{N}$ and each $(y,z) \in \mathcal{D}^{N}$, if $\{(y^\nu,z^\nu)\}$ converges to $(y,z)$, then the sequence $\{R(y^\nu,z^\nu)\}$ converges to $R(y,z)$ in $\mathbb{R}^N$.

The third axiom states that the process is additive with respect to incomes and needs. Consider the following two alternatives. In one of them we solve two redistribution problems $(y,z)$ and $(y',z')$ in two subsequent periods of time. In another, we simply address the lump sum situation at the end $(y+y',z+z')$. \emph{Additivity} guarantees that the final redistribution does not depend on the chosen alternative, so that the timing cannot alter the outcome (thus excluding arbitrariness). 

\textbf{Additivity}. For each pair $(y,z),(y',z') \in \mathcal{D}^{N}$, $R(y+y',z+z')=R(y,z)+R(y',z')$.

We also consider six additional axioms. The first three refer to natural lower bounds. They all apply under the premise that the income profile dominates the needs profile (and only for that case). And each of them formalizes a different lower bound; zero, the individual need or the average net income (after accounting for needs). Formally,

\textbf{Zero lower bound}. For each $(y,z) \in \mathcal{D}^{N}$ with $y \geq z$, and each $i \in N$, $R_{i}(y,z)\ge 0$.

\textbf{Needs lower bound}. For each $(y,z) \in \mathcal{D}^{N}$ with $y \geq z$, and each $i \in N$, $R_{i}(y,z)\ge z_{i}$.

\textbf{Net-average lower bound}. For each $(y,z) \in \mathcal{D}^{N}$ such that $y \geq z$, and each $i \in N$, $R_{i}(y,z) \ge \frac{Y-Z}{n}$.\footnote{This axiom can be rationalized via splitting the process in two: aggregate needs on the one hand, and the residual income on the other hand. The axiom then states that everyone is entitled at least an equal share of the second part.} 

Note that both \textit{needs lower bound} and \textit{net-average lower bound} imply \textit{zero lower bound}. No other logical relation exists among the three axioms. 

We also consider an individual rationality axiom stating that, for uniform needs profiles, those with larger incomes end up with higher incomes too after redistribution.\footnote{Order preservation axioms are widely used in axiomatic work. For instance, in claims problems, it is formalized stating that an agent with a larger claim than another should receive (and also lose) at least as much as the other (e.g., Thomson, 2019). Many instances appear in other related problems (e.g., Hougaard et al., 2017; Berganti\~{n}os and Moreno-Ternero, 2022; Martinez and S\'{a}nchez-Soriano, 2023).} 

\textbf{Order preservation for uniform needs}. For each $(y,z) \in \mathcal{D}^{N}$ such that $z_k=z_l$ for all $\{k,l\} \subseteq N$, and each $\{i,j\} \subseteq N$, if  $y_{i}\ge y_{j}$ then $R_i(y,z)\ge R_j(y,z)$.

Finally, we consider two axioms formalizing the principle of monotonicity, which is also widely applied in allocation problems. The first one states that, if an agent's need increases then her allocation cannot decrease. The second one is a stronger version 
adding that the above cannot be imposed at the expense of those agents with a smaller net income. 

\textbf{Need monotonicity}. For each pair $(y,z),(y,z') \in \mathcal{D}^{N}$ and each $i \in N$, if $z_i>z'_i$ and $z_j=z'_j$ for all $j \in N \backslash \{i\}$, then $R_i(y,z)\ge R_i(y,z')$.

\textbf{Strong need monotonicity}. For each pair $(y,z),(y,z') \in \mathcal{D}^{N}$ and each $i \in N$, if $z_i>z'_i$ and $z_j=z'_j$ for all $j \in N \backslash \{i\}$, then $R_i(y,z)\ge R_i(y,z')$ and $R_j(y,z)\ge R_j(y,z')$ for each $j \in N \backslash \{i\}$ such that $y_j-z_j \leq y_i-z_i$.

We conclude this section presenting some natural rules that obey most of the previous axioms. 

First, we introduce the rule that leaves incomes untouched. 

\textbf{Laissez Faire} ($R^{L}$). For each $(y,z) \in \mathcal{D}^{N}$ and each $i \in N$, 
$$R^{L}_i(y,z)=y_{i}.$$

Second, we introduce its polar rule, which imposes full redistribution. 

\textbf{Full Redistribution} ($R^{F}$). For each $(y,z) \in \mathcal{D}^{N}$ and each $i \in N$, 
$$R^{F}_i(y,z)=\frac{Y}{n}.$$

Third, we introduce the rule that assigns needs first, and then reallocates equally the residual.

\textbf{Need-Adjusted Full Redistribution} ($R^{A}$). For each $(y,z) \in \mathcal{D}^{N}$ and each $i \in N$, 
$$R^{A}_i(y,z)= z_i + \frac{Y-Z}{n}.$$

As we show in the next section, these three rules essentially generate all the rules that satisfy the previous axioms.

\section{Characterization results}
Our main result characterizes all the rules that satisfy our three basic axioms (\textit{equal treatment of equals}, \textit{continuity}, and \textit{additivity}). They are precisely linear combinations of the three  rules introduced above. Note that, as the statement indicates, the parameters for the linear combinations ($\lambda_1$ and $\lambda_2$) can be negative.

\begin{theorem}\label{ADD_AN}
A rule satisfies equal treatment of equals, continuity, and additivity if and only if there exist $(\lambda_1,\lambda_2) \in \mathbb{R}^2$ such that, for each $(y,z) \in \mathcal{D}^N$,
$$
R(y,z) = \lambda_1 R^{L}(y,z) + \lambda_2 R^{F}(y,z) + (1 - \lambda_1 - \lambda_2) R^{A}(y,z).
$$

\end{theorem}

Theorem 1 states that all the impartial, continuous and additive rules can be decomposed in three parts. One part refers to laissez faire; another to (gross) full redistribution; and yet a third to net full redistribution, after accounting for needs. We next show that the decomposition can be more specific after adding some additional axioms. More precisely, we can characterize subfamilies of the general family characterized in Theorem \ref{ADD_AN} adding the auxiliary axioms we introduced above. And it turns out that adding these axioms allows us to dismiss the axiom of continuity. 
These subfamilies are actually made of compromises (via convex combinations) among each of the resulting pairs made from the three rules mentioned above. 

\begin{theorem}\label{ADD_AN_THIRD}
Let $R$ be a rule satisfying equal treatment of equals and additivity.
\begin{itemize}
  \item[(a)] $R$ also satisfies zero lower bound, order preservation for uniform needs, and strong need monotonicity if and only if there exists $\delta_1 \in [0,1]$ such that, for each $(y,z) \in \mathcal{D}^N$,
  $$
  R(y,z) = \delta_1 R^{L}(y,z) + (1-\delta_1) R^{F}(y,z).
  $$
  \item[(b)] $R$ also satisfies needs lower bound and order preservation for uniform needs if and only if there exists $\delta_2 \in [0,1]$ such that, for each $(y,z) \in \mathcal{D}^N$,
  $$
  R(y,z) = \delta_2 R^{L}(y,z) + (1-\delta_2) R^{A}(y,z).
  $$
  \item[(c)] $R$ also satisfies net-average lower bound and need monotonicity if and only if there exists $\delta_3 \in [0,1]$ such that, for each $(y,z) \in \mathcal{D}^N$,
  $$
  R(y,z) = \delta_3 R^{F}(y,z) + (1-\delta_3) R^{A}(y,z).
  $$
\end{itemize}

\end{theorem}

We conclude our axiomatic analysis providing characterizations of our three focal rules, by means of various combinations of the axioms introduced above.

\begin{theorem}\label{char_focal}
  The following statements hold:
  \begin{itemize}
      \item[(a)] $R$ satisfies additivity, needs lower bound and strong needs monotonicity if and only if it is the laissez-faire rule.
      \item[(b)] $R$ satisfies additivity, net-average lower bound and strong needs monotonicity if and only if it is the full redistribution rule.
      \item[(c)] $R$ satisfies additivity, needs lower bound, and net-average lower bound if and only if it is the need-adjusted full redistribution rule.
  \end{itemize}
  \end{theorem}

\section{Decentralization}
We have provided  
normative foundations for families of allocation rules. The axiomatic analysis can be pursued further to single out a specific rule within the families (as exemplified in Theorem \ref{char_focal}). Alternatively, we could explore such a problem differently, taking a decentralized approach, as we do in this section. More precisely, we
study whether the choice of a rule within a family could be made by means
of simple majority voting, letting each individual vote for a rule within the family. Due to the overwhelming existence of majority cycles, one should normally
not expect a positive answer to this question. Somewhat surprisingly, we do obtain partially positive answers in our setting.

We first provide some formal definitions. Given a problem $(y,z) \in \mathcal{D}^N$, we say that $R(y,z) $ is a \emph{majority winner} (within the set of rules $\mathcal{R}$, considered as the domain for voting) for $(y,z)$ if there is no other rule $R^{\prime }\in \mathcal{R}$ such that $R_{i}^{\prime }(y,z)>R_{i}(y,z)$ for a majority of agents. We say that the family of rules $\mathcal{R}$ has a \emph{majority voting equilibrium} if there is at least one majority winner (within $\mathcal{R}$, considered as the domain for voting) for each problem $(y,z)\in \mathcal{D}^N$. 

Let $\left\{ R^{\delta_1}\right\} _{\delta_1 \in [0,1]}$, $\left\{ R^{\delta_2}\right\} _{\delta_2 \in [0,1]}$, and $\left\{ R^{\delta_3}\right\} _{\delta_3 \in [0,1]}$, be the three families introduced above. Formally, for each $(y,z) \in \mathcal{D}^N$, 
$$R^{\delta_1}(y,z) = \delta_1 R^{L}(y,z) + (1-\delta_1) R^{F}(y,z),$$ 
$$R^{\delta_2}(y,z) = \delta_2 R^{L}(y,z) + (1-\delta_2) R^{A}(y,z),$$ 
$$R^{\delta_3}(y,z) = \delta_3 R^{F}(y,z) + (1-\delta_3) R^{A}(y,z).$$

For each $(y,z)\in \mathcal{D}^N$, we consider the following
partition of $N$, with respect to the average income ($\frac{Y}{n}$): $%
N^{y}_{l}(y,z) =\{i\in N:y_{i}<\frac{Y}{n}\}$, $N^{y}_{u}(y,z) =\{i\in N:y_{i}>\frac{Y}{n}\}$,\ and $N^{y}_{e}(y,z)
=\{i\in N:y _{i}=\frac{Y}{n}\}$. That is, taking the average (pre-tax) income as the benchmark threshold, we consider three groups
referring to agents with incomes below, above, or exactly at, the threshold. For ease of notation, we simply write $N^{y}_{l}$, $N^{y}_{u},$ and $N^{y}_{e}$. Note that $n=|N^{y}_{l}|+|N^{y}_{u}|+|N^{y}_{e}|$.

\begin{proposition}
\label{Majority 1} Let $\left\{ R^{\delta_1}\right\} _{\delta_1 \in [0,1]}$ be the domain of rules among which agents can vote.
For each $(y,z)\in \mathcal{D}^N$, the following statements hold:

$\left( i\right) $ If $2|N^{y}_{l}|>n$, then $R^{F}(y,z) $ is the unique majority winner.

$\left( ii\right) $ If $2|N^{y}_{u}|>n$, then $R^{L}(y,z)$ is the unique majority winner.

$\left( iii\right) $ Otherwise, 
each $R^{\delta_1}(y,z) $ is a majority winner.
\end{proposition}

For each $(y,z)\in \mathcal{D}^N$, we can also consider the following
partition of $N$, with respect to the average net income ($\frac{Y-Z}{n}$): $%
N_{l}(y,z) =\{i\in N:y_{i}-z_{i}<\frac{Y-Z}{n}\}$, $N_{u}(y,z) =\{i\in N:y_{i}-z_{i}>\frac{Y-Z}{n}\}$,\ and $N_{e}(y,z)
=\{i\in N:y_{i}-z_{i}=\frac{Y-Z}{n}\}$. That is, taking the average net income as the benchmark threshold for net incomes, we consider three groups
referring to agents below, above, or exactly at, the threshold. For ease of exposition, we simply write $N_{l}$, $N_{u},$ and $N_{e}$.
We then have the following result.

\begin{proposition}
\label{Majority 2} Let $\left\{ R^{\delta_2}\right\} _{\delta_2 \in [0,1]}$ be the be the domain of rules among which agents can vote. 
For each $(y,z)\in \mathcal{D}^N$, the following statements hold:

$\left( i\right) $ If $2|N_{l}|>n$, then $R^{A}(y,z) $ is the unique majority winner.

$\left( ii\right) $ If $2|N_{u}|>n$, then $R^{L}(y,z)$ is the unique majority winner.

$\left( iii\right) $ Otherwise, 
each $R^{\delta_2}(y,z) $ is a majority winner.
\end{proposition}

Finally, for each $(y,z)\in \mathcal{D}^N$, we can also consider the following partition of $N$, with respect to the average need ($\frac{Z}{n}$): $%
N^{z}_{l}(y,z) =\{i\in N:z_{i}<\frac{Z}{n}\}$, $N^{z}_{u}(y,z) =\{i\in N:z_{i}>\frac{Z}{n}\}$, and $N^{z}_{e}(y,z)
=\{i\in N:z _{i}=\frac{Z}{n}\}$. That is, taking the average need as the benchmark threshold (for needs), we consider three groups
referring to agents with needs below, above, or exactly at, the threshold. For ease of exposition, we simply write $N^{z}_{l}$, $N^{z}_{u},$ and $N^{z}_{e}$.
We then have the following result.

\begin{proposition}
\label{Majority 3} Let $\left\{ R^{\delta_3}\right\} _{\delta_3 \in [0,1]}$ be the domain of rules among which agents can vote. 
For each $(y,z)\in \mathcal{D}^N$, the following statements hold:

$\left( i\right) $ If $2|N^{z}_{l}|>n$, then $R^{F}(y,z) $ is the unique majority winner.

$\left( ii\right) $ If $2|N^{z}_{u}|>n$, then $R^{A}(y,z)$ is the unique majority winner.

$\left( iii\right) $ Otherwise, 
each $R^{\delta_3}(y,z) $ is a majority winner.
\end{proposition}


\section{Lorenz dominance}
In this section, we uncover the structure of the families of rules considered above. To do so, we introduce some notation first. For each $x \in \mathbb{R}^N$, we define the partial sum of order $k \in \{1,\ldots,n\}$ as $S^k(x) = \sum_{i=1}^k x_{(i)}$, where $x_{(i)}$ is the $i$-th lowest coordinate of $x$, i.e., $x_{(1)}\leq x_{(2)}\leq...\leq x_{(n)}$. For each pair $x,\overline{x} \in \mathbb{R}^N$, such that $S^n(x) = S^n(\overline{x})$, we say that $x$ dominates $\overline{x}$ in the sense of Lorenz, which we write as $x \succeq_L \overline{x}$, if and only if $S^k(x) \ge S^k(\overline{x}) \; \; \forall k \in \{1,\ldots,n-1\}$. When $x$ dominates $\overline{x}$ in the sense of Lorenz, one can state that $x$ is unambiguously \textquotedblleft more egalitarian\textquotedblright\ than $\overline{x}$.

In our setting, we say that a rule $R$ \textit{Lorenz dominates} another rule $\overline{R}$ if, for each $(y,z) \in \mathcal{D}^N$, $R(y,z)$ dominates $\overline{R}(y,z)$ in the sense of Lorenz, which we write as $R \succeq_{L} \overline{R}$.\footnote{Our analysis thus refers to the resulting (one dimensional) post-income distributions. A possible analysis of Lorenz dominance extended to needs could be carried out following the proposals in Atkinson and Bourguignon (1987), Bourguignon (1989) or, more recently, Faure and Gravel (2021).} As the Lorenz criterion is a partial ordering, one might not expect to be able to perform many comparisons of vectors. It turns out, however, that two of our families of rules can be fully ranked according to this criterion (as stated in the next result).

\begin{proposition}\label{Lorenz 1}
For each $\delta \in [0,1]$,  
\begin{itemize}
    \item[(a)] Let $R^\delta =\delta R^L + (1-\delta) R^F$. If $0 \le \delta \le \overline{\delta}\leq 1$ then $R^\delta \succeq_{L} R^{\overline{\delta}}$.
    \item[(b)] Let $R^\delta =\delta R^F + (1-\delta) R^{A}$. If $0 \le \delta \le \overline{\delta} \leq 1$ then $R^{\overline{\delta}} \succeq_{L} R^\delta$. 
\end{itemize}

\end{proposition}

The previous result indicates that the compromises between the laissez faire and the full redistribution rules, as well as the compromises between the full redistribution and need-adjusted full redistribution rules are fully ranked according to the Lorenz criterion. The parameter defining each family of compromises can actually be interpreted as an index of the egalitarianism rules within each family convey.  On the other hand, as the next result states, the compromises between the laissez faire and the need-adjusted full redistribution rules are not ordered according to the Lorenz domination. 

\begin{proposition}\label{Lorenz 2}
The family of rules described by
$$
R^\delta =\delta R^L + (1-\delta) R^{A},
$$
where $\delta \in [0,1]$, cannot be fully ranked according to the Lorenz criterion.
\end{proposition}

Propositions \ref{Lorenz 1} and \ref{Lorenz 2} imply that some (but not all) pairs of rules within the family characterized in Theorem \ref{ADD_AN} can be ranked according to the Lorenz criterion.

\section{Optimal taxation}
In this section, we connect our work with the literature on optimal income taxation.  

In practice, income redistribution takes place through income taxes and subsidies chosen by the government. We can naturally define the taxation rule associated to a redistribution rule in our setting as follows. For each $(y,z) \in \mathcal{D}^{N}$, each $i\in N$, and each allocation rule $R$, let
$$
T_i(y,z)=y_i-R_i(y,z). 
$$
We can also naturally translate the axioms in Theorem \ref{ADD_AN} for the redistribution rule $R$ into axioms for the taxation rule $T$, not only technically but also conceptually. To wit, \textit{equal treatment of equals} requires that equal agents pay equal taxes, \textit{continuity} guarantees that minor variations in the input variables do not significantly affect taxes, and \textit{additivity} states that each agent must pay the same amount of tax as in the lump sum situation $(y + y', z + z')$, even if the timing of tax payment is divided into the earlier situation $(y, z)$ and the later situation $(y', z')$.

Theorem 1 can then be written as follows:

\begin{theorem}\label{tax}
    A (taxation) rule $T$ satisfies equal treatment of equals, continuity, and additivity if and only if there exist $(\lambda_1,\lambda_2) \in \mathbb{R}^2$ such that, for each $(y,z) \in \mathcal{D}^N$,
    $$
    T_i(y,z) = (1-\lambda_1) \left(y_i-\frac{Y}{n}\right) -(1-\lambda_1-\lambda_2) \left(z_i-\frac{Z}{n}\right).
    $$
\end{theorem}
Theorem \ref{tax} provides an axiomatic justification for a linear income tax system. All the elements in which this system decomposes have a clear and meaningful interpretation. Namely, $(1-\lambda_1)$ is a marginal income tax rate, $\frac{Y}{n}$ is a tax deduction, $(1-\lambda_1-\lambda_2)$ is a marginal subsidy rate for need, and $(1-\lambda_1-\lambda_2) \left(z_i-\frac{Z}{n}\right)$ means lump-sum subsidy for need.

The above connects to the \textit{linear income tax system} that appears in theoretical models of optimal income taxation in public economics (e.g., Dixit and Sandmo, 1977; Ihori, 1987). 
One of the fundamental questions concerning income redistribution that has been discussed in public economics is whether the optimal (marginal) income tax rate, with income taxation limited to a linear system, is greater as social concern for inequality increases. According to the literature, the answer to this question is positive (e.g., Atkinson and Stiglitz, 1980). This is confirmed by our analysis. More precisely, the translation of our Proposition \ref{Lorenz 1} above to this setting implies that as the marginal income tax rate increases, the degree of inequality (in the sense of the Lorenz criterion) in the family of rules decreases. 

\section{Empirical illustration}

In this section, we illustrate our previous normative analysis. To do so, we resort to Eurostat's Household Budget Survey (HBS), 
which provides information on households' expenditure on goods and services at the national level
across the European Union. We use the microdata from the last available wave (2015). This survey consists of 271748 households and more than 400 variables concerning their social and demographic structure, income, and expenditure. As for the latter, we have information on how much each household expends on consuming different items: food (at, or away from home), housing (including shelter, utilities, fuels, furnishing, or housekeeping supplies), apparel and services, transportation (vehicle purchases, gasoline, maintenance, or public transportation), health care, personal care, and education, among others.

\begin{figure}[h]
  \centering
  \includegraphics[scale=1]{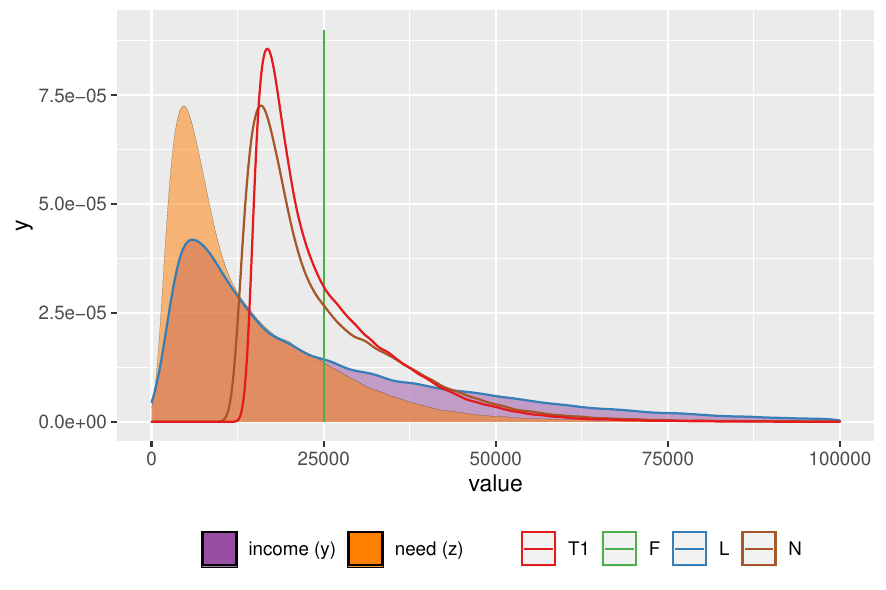}
  \caption{Distributions of incomes, needs, and the reallocations resulting from the rules. \label{fig1}}
\end{figure}

In our theoretical model we have two variables: income ($y$) and need ($z$). We identify $y$ with the net income variable in the HBS. We construct $z$ as the total consumption expenditure minus the expenditure in those items we do not classify as ``need". Although we acknowledge that categorizing some consumption items as needs may be a subjective decision, the illustrations we present in this section would not differ significantly altering this decision to some extent.\footnote{The complete list of items in the HBS is available at the Eurostat's webpage (https://ec.europa.eu/eurostat/web/microdata/household-budget-survey). From all those, we consider the following as need: food, non-alcoholic beverages, clothing and footwear, housing (water, electricity, gas and other fuels), furniture and furnishings, household textiles, household appliances, glassware, tableware and household utensils, health, purchase of second-hand motor-cars, operation of personal transport equipment, transport services, communication, and education.}

In Figure \ref{fig1}, we plot the distributions of income and needs of the European households. The average income and need are 25053.8 EUR/year and 13830.5 EUR/year, respectively. 79.8\% of households have incomes below 40000 EUR, and 96.2\% of households have needs below that threshold. Besides, 10.7\% of the European households do not have enough income to afford their own needs. As for the rules, we have represented the distributions corresponding to our three fundamental rules; namely, laissez faire ($R^L$), full redistribution ($R^F$), and need-adjusted full redistribution ($R^{A}$). Obviously, laissez faire mimics the income distribution. With full redistribution, all households would obtain the same amount (the mean income). More interestingly, the need-adjusted full redistribution shifts the distribution of needs by an amount that depends on the average net income. More precisely, the rule guarantees that any household gets, at least, 11232.5 EUR (the average net income), which can be interpreted as a basic (household) income. Finally, we also plot a member of the family of rules characterized in Theorem \ref{ADD_AN}, compromising among the three fundamental rules, for the parameter values $\lambda_1=\frac{3}{10}$ and $\lambda_2=\frac{4}{10}$.

\begin{figure}[h]
  \centering
  \includegraphics[scale=0.65]{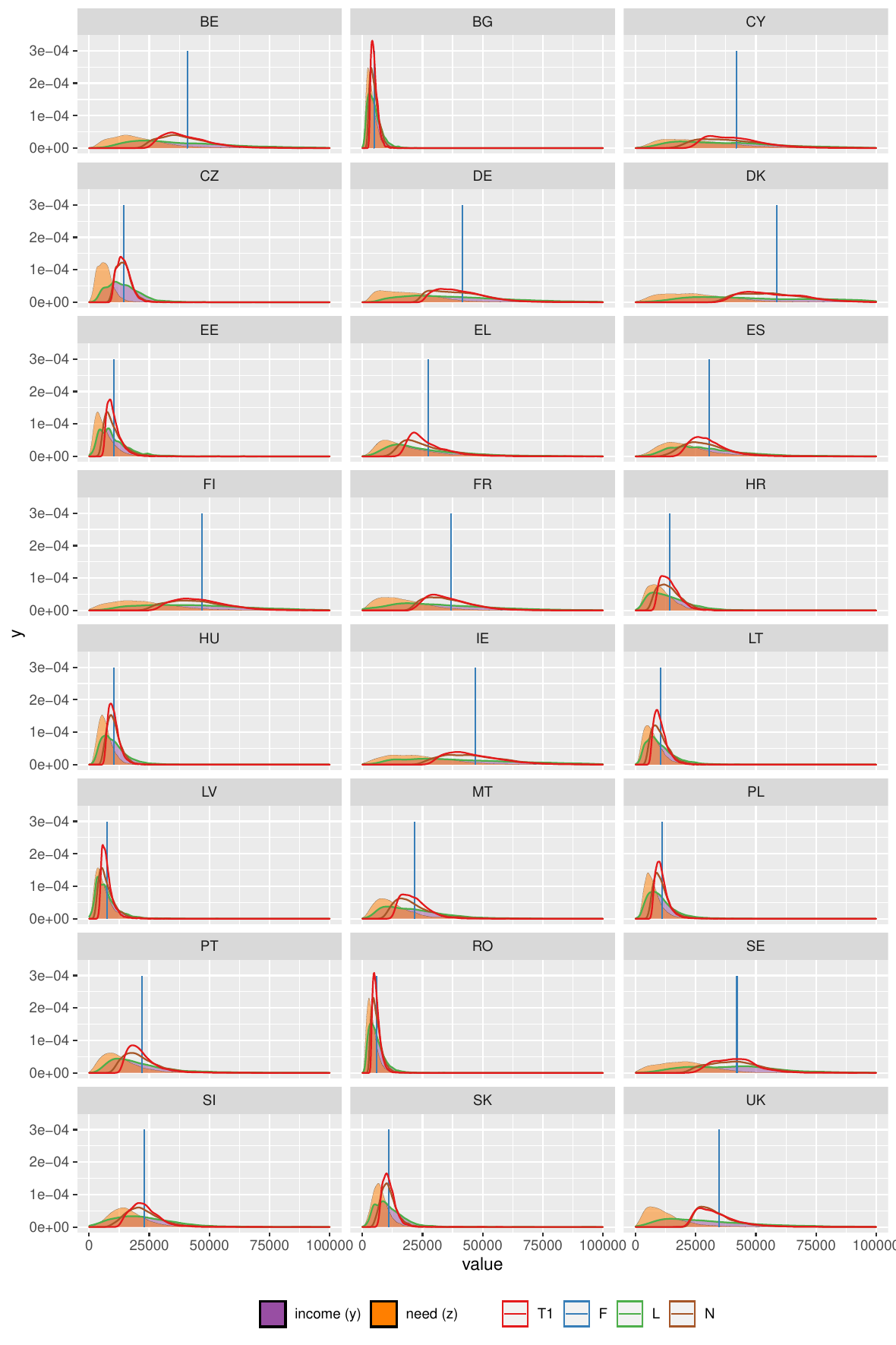}
  \caption{Distributions by countries. \label{fig2}}
\end{figure}

In Figure \ref{fig2} we represent the incomes, needs, and redistributions at the national level. Not surprisingly, different countries exhibit different patterns.\footnote{The countries in Figure \ref{fig2} (in the alphabetical order of their abbreviations therein) are the following: Belgium (BE), Bulgaria (BG), Cyprus (CY), Czech Republic (CZ), Germany (DE), Denmark (DK), Estonia (EE), Greece (EL), Spain (ES), Finland (FI), France (FR), Croatia (HR), Hungary (HU), Ireland (IE), Lithuania (LT), Latvia (LV), Malta (MT), Poland (PL), Portugal (PT), Romania (RE), Sweden (SE), Slovenia (SI), Slovakia (SK), and United Kingdom (UK).} For some nations (e.g., Bulgaria or Romania) the four proposals result in rather similar redistributions. For some other countries (e.g. Malta, Portugal, or Greece) the differences are more significant. As we can observe, in the case of the United Kingdom the compromise rule we selected almost replicates the distribution obtained from the need-adjusted full redistribution rule. Typically, the differences between these two rules are more evident. For instance, for Lithuania or Hungary, the means of both rules are very close, but the distributions are more concentrated around the corresponding mean in the case of the compromise rule. In the cases of Cyprus or Germany, the compromise rule appears to be a right shift of the need-adjusted full redistribution rule.

\begin{center}
Insert Figure 2 about here
\end{center}

We conclude this section mentioning that our empirical illustration might suggest to examine econometrically whether the redistribution of income by each country's progressive income tax system corresponds to the consequences of the rule satisfying social fairness as described in each theorem. In other words, to estimate $\lambda$ such that the consequence of the rule $R^{\lambda}(y,z)$ corresponds to the real income redistribution through taxation and subsidies. Unfortunately, the Household Budget Survey does not provide all the necessary data for this exercise; namely, both the gross and net income of each household (or, equivalently, one of them and the corresponding taxes they pay from which we could infer the other). 
  
\section{Discussion}
We have presented in this paper an axiomatic approach to the problem of (income) redistribution with needs. We have provided normative foundations for (redistribution) rules that arise from (linear) combinations of three fundamental pillars: laissez faire and full redistribution, with or without adjusting first for the assessment of needs. Specific compromises among each of those two pillars (via convex combinations) have also been scrutinized in our analysis. First, via further characterizations. Second, via majority voting. Third, via the Lorenz criterion.

Our results on majority voting provide us with the specific equilibrium within each family of rules we highlight. These equilibria happen to be one of the extreme members within each family, depending on the skewness of the income or need distributions in each case. These results are reminiscent to existing results in the literature. For instance, Corch\'{o}n and Puy (1998) study the problem in which a group of individuals owns collectively a technology which produces a consumption good from an input. They characterize the family of sharing rules (associating input contributions with a vector of consumption) that satisfy Pareto efficiency and individual rationality, which happen to be a convex combination of two basic rules. As in our case, the outcome of majority voting on this family of sharing rules is one of those basic rules depending on the skewness of the distribution of labor contributions. Similarly, Berganti\~{n}os and Moreno-Ternero (2020, 2021, 2023) study the so-called broadcasting problems that arise when revenues raised (collectively) from broadcasting sports leagues have to be shared among participating clubs. They show that the family of rules compromising between two basic rules are characterized by the axioms of \textit{equal treatment of equals}, \textit{additivity} and a reasonable \textit{upper bound}. If clubs are allowed to vote among members of that family, the outcome of majority voting is one of those basic rules depending on the skewness of the distribution of overall broadcasting audiences.

The existence of majority voting equilibria (although not their specific description) could also be obtained as a consequence of the so-called \textit{single-crossing property} these families exhibit. More precisely, in each of the families, one can separate those agents that benefit from the application of one rule or another (within the family), depending on the rank of their incomes, needs, or adjusted incomes. It is well known that this property guarantees that the social preference relationship obtained under majority voting is transitive, and corresponds to the median voter's (e.g., Gans and Smart, 1996). 
Another consequence of the single-crossing property is that it guarantees progressivity comparisons of schedules (e.g., Jakobsson, 1976; Hemming and Keen, 1983). Thus, the property could also help partially to obtain some of our results on Lorenz rankings.

Our analysis of the decentralized process assumes that voters select separately from each of the three combination-of-two-focal-rules subfamilies. One might find more interesting to offer voters to choose among all the rules in the whole family compromising among the three focal rules (which we characterize in our main result). But such a family is \textit{too large} to allow for the existence of majority winners. That is, for each problem within a large class, and each solution for it (obtained from one of the rules in the large family), there exists another solution (obtained from another rule within the family) that is strictly preferred by a majority of agents. Such a negative result is another instance of Condorcet's paradox of voting, which is perhaps best exemplified by the problem of determining the division of a cake by majority rule (Hamada 1973). 

We have also explored the implications of our results in the context of taxation. In particular, we have provided an axiomatic justification for a linear income tax system, via a family of admissible options. We have also seen that, as the marginal income tax rate increases within this family, the degree of inequality decreases, which confirms the conventional wisdom within the public economics literature.

We acknowledge that a somewhat controversial aspect of our analysis is the modelling of needs. There are compelling reasons to defend that needs might be multidimensional, subjective or qualitative concepts. Thus, they might not easily be reduced to a number. Nevertheless, a rationale for doing so may come from the well-known capabilities approach (e.g., Nussbaum and Sen, 1993; Sen, 1999), where needs could be interpreted as monetized requirements to satisfy the functioning of agents. Likewise, the so-called Quality-Adjusted-Life Years, Healthy Years Equivalent, or Disability-Adjusted-Life Years are standard options in the economic evaluation of health care programs to reduce individual health statuses into a number (e.g., Hougaard et al., 2013; Moreno-Ternero and \O sterdal, 2017; 
Moreno-Ternero et al., 2023).

Another aspect of our analysis that deserves further scrutiny is the axiom of \textit{additivity}. As mentioned above it has a long tradition of use, that can be traced back to Shapley (1953).\footnote{The notion is also used in the context of bankruptcy problems by many authors. For instance, additivity with respect to one variable is used in Chun (1988) and Moulin (1987), and with respect to two variables by Berganti\~ nos and Vidal-Puga (2004).}
 Its normative content is framed within the broad category of axioms that formalize robustness to the choice of perspective in evaluating a change, from which a variety of invariance requirements can be derived.\footnote{This category includes other axioms with a long tradition in axiomatic work. such as composition properties, separability or consistency (e.g., Thomson, 2011, 2019, 2023).} 
 A natural application of \textit{additivity} occurs when we face allocations in different periods of time. In our setting, a problem might be faced, say, at the end of the year or at the end of its two semesters. What the axiom then indicates is that the solution to the former should coincide with the aggregation of the solutions to the other two. Finally, the appeal of the axiom can also be perfectly illustrated in the following example, borrowed from Thomson (2023). Suppose a building contractor wins two contracts for renovation projects. 
The contractor hires the usual team to work on both projects and could therefore calculate what to pay the team workers project by project and award to each their compensation for each project separately. Another perspective is to think of the two projects as one big project and calculate their compensations only once. The axiom of \textit{additivity} requires that team workers should get the same total compensation either way. Otherwise, people would be affected differently depending upon the perspective the contractor would take, which would bring arbitrariness to the process. In that sense, \textit{additivity} can be seen as fairness axiom, as it excludes arbitrariness.\footnote{Note that, whereas in this example there is only one team to take care of two projects and, thus, the notion requires a rule to be additive with respect to the projects, in our setting the notion requires a rule to be additive with respect to income and needs.} 

To conclude, we mention that we have also provided an illustration of our analysis resorting to the EU Household Budget Survey. Based on it, we obtain the distributions of income and needs of European households and represent the patterns that different nations exhibit with respect to the redistribution rules we characterize in our analysis. We have shown that different countries exhibit different redistribution patterns. For some nations, the three pillars (and, thus, compromises among them) result in rather similar redistributions. For some other countries, differences are more significant.


\section*{Appendix}
\subsection*{Proof of Theorem \ref{ADD_AN}}
    It is not difficult to show that all rules within the family satisfy the axioms in the statement. Conversely, let $R$ be a rule satisfying \textit{equal treatment of equals}, \emph{continuity}, and \textit{additivity}. Let $(y,z) \in \mathcal{D}^N$. By \emph{additivity},
    $$
    R(y,z) = R(y-z,0) + R(z,z).
    $$
    Let $i \in N$. By \emph{additivity} and \emph{continuity} (e.g., Eichorn, 1978; Acz\'{e}l, 2006),
    $$
    R_i(y-z,0)=\sum_{k=1}^{n}  R_i((y_{k}-z_{k},0_{-k}),0)=\sum_{k=1}^{n}  (y_{k}-z_{k}) R_i((1_k,0_{-k}),0),
    $$
    where, for each $k\in N$, $0_{-k}$ denotes zero entries for all coordinates in $N\setminus\{k\}$, and $1_k$ denotes a $1$ entry in the $k$-th coordinate.  
    
    Now, by \emph{equal treatment of equals}, there exists $(\alpha_1, \dots \alpha_n) \in \mathbb{R}^n$ such that
    $$
    R_i((1_k,0_{-k}),0) =
    \begin{cases}
    \alpha_k & \text{if } i=k \\
    \frac{1-\alpha_k}{n-1} &\text{if } i \neq k
    \end{cases}.
    $$
    We show now that $\alpha_1 = \dots = \alpha_n= \alpha$. To do that, let $k \in N \backslash \{1\}$. Then, $R_1((1_1,0_{-1}),0)=\alpha_1$, $R_k((1_1,0_{-1}),0)=\frac{1-\alpha_1}{n-1}$, $R_1((1_k,0_{-k}),0)=\frac{1-\alpha_k}{n-1}$, and $R_k((1_k,0_{-k}),0)=\alpha_k$. By \emph{additivity},
    $$
    R_1((1_{\{1,k\}},0_{-\{1,k\}}),0) = R_1((1_1,0_{-1}),0)+R_1((1_k,0_{-k}),0)=\alpha_1 + \frac{1-\alpha_k}{n-1},
    $$
    and
    $$
    R_k((1_{\{1,k\}},0_{-\{1,k\}}),0) = R_k((1_1,0_{-1}),0)+R_k((1_k,0_{-k}),0) = \frac{1-\alpha_1}{n-1} + \alpha_k,
    $$
    where, for each $k\in N$, $0_{-\{1,k\}}$ denotes zero entries for all coordinates in $N\setminus\{1,k\}$, and $1_{\{1,k\}}$ denotes a $1$ entry in the first and $k$-th coordinates. As agents $1$ and $k$ have equal incomes and equal needs in the problem $((1_{\{1,k\}},0_{-\{1,k\}}),0)$, it follows from \emph{equal treatment of equals} that
    $$
    \alpha_1 + \frac{1-\alpha_k}{n-1} = \frac{1-\alpha_1}{n-1} + \alpha_k.
    $$
    Thus, $\alpha_1 = \dots = \alpha_n= \alpha$. Therefore,
    $$
    R_i(y-z,0)=\alpha(y_i-z_i) + \frac{1-\alpha}{n-1} \sum_{k \neq i}(y_k-z_k) = \alpha(y_i-z_i) + \frac{1-\alpha}{n-1} \left( Y-Z-(y_i-z_i) \right).
    $$
    Similarly, by \emph{additivity}, we know that
    $$
    R_i(z,z) =\sum_{k=1}^{n}  z_{k} R_i((1_k,0_{-k}),(1_k,0_{-k})).
    $$
    By \emph{equal treatment of equals}, there exists $(\beta_1, \dots \beta_n) \in \mathbb{R}^n$ such that
    $$
    R_i((1_k,0_{-k}),(1_k,0_{-k})) =
    \begin{cases}
    \beta_k & \text{if } i=k \\
    \frac{1-\beta_k}{n-1} & i \neq k
    \end{cases}.
    $$
    As before, 
    we obtain that $\beta_1 = \ldots = \beta_k = \beta$. To wit, let $k \in N \backslash \{1\}$. Then, $R_1((1_1,0_{-1}),(1_1,0_{-1}))=\beta_1$, $R_k((1_1,0_{-1}),(1_1,0_{-1}))=\frac{1-\beta_1}{n-1}$, $R_1((1_k,0_{-k}),(1_k,0_{-k}))=\frac{1-\beta_k}{n-1}$, and $R_k((1_k,0_{-k}),(1_k,0_{-k}))=\beta_k$. By \emph{additivity},
    \begin{align*}
    R_1((1_{\{1,k\}},0_{-\{1,k\}}),(1_{\{1,k\}},0_{-\{1,k\}})) &= R_1((1_1,0_{-1}),(1_1,0_{-1}))+R_1((1_k,0_{-k}),(1_k,0_{-k})) \\
    &= \beta_1 + \frac{1-\beta_k}{n-1},
    \end{align*}
    and
    \begin{align*}
    R_k((1_{\{1,k\}},0_{-\{1,k\}}),(1_{\{1,k\}},0_{-\{1,k\}})) &= R_k((1_1,0_{-1}),(1_1,0_{-1}))+R_k((1_k,0_{-k}),(1_k,0_{-k})) \\
    &= \frac{1-\beta_1}{n-1} + \beta_k.
    \end{align*}
    As  agents $1$ and $k$ have equal incomes and equal needs in the problem $((1_{\{1,k\}},0_{-\{1,k\}}),(1_{\{1,k\}},0_{-\{1,k\}}))$, it follows from \emph{equal treatment of equals} that
    $$
    \beta_1 + \frac{1-\beta_k}{n-1} = \frac{1-\beta_1}{n-1} + y_k.
    $$
    That is, $\beta_1 = \dots = \beta_n= \beta$. Therefore,
    $$
    R_i(z,z)=\beta z_i + \frac{1-\beta}{n-1} \sum_{k \neq i} z_k = \beta z_i + \frac{1-\beta}{n-1} \left( Z-z_i \right).
    $$
    Altogether,
    $$
    R_i(y,z) = \alpha(y_i-z_i) + \frac{1-\alpha}{n-1} \left( Y-Z-(y_i-z_i) \right) + \beta z_i + \frac{1-\beta}{n-1} \left( Z-z_i \right),
    $$
    which can be re-written as
    $$
    R_i(y,z) = \frac{n\alpha-1}{n-1} y_i + \frac{n(1-\beta)}{n-1} \frac{Y}{n}+ \frac{n(\beta-\alpha)}{n-1} \left[ z_i + \frac{Y-Z}{n} \right].
    $$
    If we define $\lambda_1=\frac{n\alpha-1}{n-1}$, $\lambda_2=\frac{n(1-\beta)}{n-1}$, and $\lambda_3=\frac{n(\beta-\alpha)}{n-1}=1 - \lambda_1 - \lambda_2$, we obtain that
    $$
    R(y,z) = \lambda_1 R^L(y,z) + \lambda_2 R^F(y,z) + (1 - \lambda_1 - \lambda_2) R^{A}(y,z),
    $$
    As $y \in \mathbb{R}$ and $\beta \in \mathbb{R}$, it follows that $\lambda_1 \in \mathbb{R}$ and $\lambda_2 \in \mathbb{R}$, which concludes the proof. \qed
 
\subsection*{Proof of Theorem \ref{ADD_AN_THIRD}}
We prove first the straightforward implications of each statement. Theorem \ref{ADD_AN} guarantees that all rules in the statements satisfy \textit{equal treatment of equals} and \textit{additivity}. We then shift towards the remaining axioms.
\begin{itemize}
    \item[(a)] As for \emph{zero lower bound}, both laissez faire and the full redistribution rules obviously satisfy it. Then, all convex combinations of them do so too. 
    Notice that none of those rules depends on the vector of needs. Therefore, \emph{strong need monotonicity} is trivially satisfied too. Finally, as for \emph{order preservation for uniform needs}, let $(y,z) \in \mathcal{D}^{N}$ be such that $z_k=z_l$ for all $\{k,l\} \subseteq N$, and let $\{i,j\} \subseteq N$ such that $y_{i}\ge y_{j}$. Then, for each $\delta_1 \in [0,1]$,
    $$
    R_i(y,z) = \delta_1 y_i + (1-\delta_1) \frac{Y}{n} \geq \delta_1 y_j + (1-\delta_1) \frac{Y}{n} = R_j(y,z),
    $$
    as desired.

    \item[(b)] As for \textit{needs lower bound}, let $(y,z) \in \mathcal{D}^N$ be such that $y \geq z$, and let $i \in N$. We need to show that, for each $\delta_2 \in [0,1]$,
    $$
    \delta_2 y_i + (1-\delta_2) \left[ z_i + \frac{Y-Z}{n}\right] \ge z_i.
    $$
    Or, equivalently, that
    $$
    \delta_2 \left[y_i-z_i-\frac{Y-Z}{n} \right] \ge - \frac{Y-Z}{n}.
    $$
    which follows from the fact that $y_j \ge z_j$ for each $j \in N$. 

    As for \emph{order preservation for uniform needs}, let $(y,z) \in \mathcal{D}^{N}$ be such that $z_k=z_l$ for all $\{k,l\} \subseteq N$, and let $\{i,j\} \subseteq N$ be such that $y_{i}\ge y_{j}$. Then, for each $\delta_2 \in [0,1]$,
    $$
    R_i(y,z)=\delta_2 y_i + (1-\delta_2)\left[z_i + \frac{Y-Z}{n} \right] \ge \delta_2 y_j + (1-\delta_2)\left[ z_j + \frac{Y-Z}{n} \right] = R_j(y,z),
    $$
as desired. 
    \item[(c)] 
    As for \textit{net-average lower bound}, let $(y,z) \in \mathcal{D}^N$ be such that $y \geq z$, and let $i \in N$. We need to show that, for each $\delta_3 \in [0,1]$,
    $$
   \delta_3 \frac{Y}{n} + (1-\delta_3) \left[ z_i + \frac{Y-Z}{n}\right] \ge \frac{Y-Z}{n}.
    $$
    Or, equivalently, that
    $$
    \delta_3 \left[ z_i - \frac{Z}{n} \right] \le z_i.
    $$
    As $\delta_3\ge 0$ and $z_i\ge 0$, the condition trivially holds when $z_i - \frac{Z}{n} \le 0$. Assume then that $z_i - \frac{Z}{n} \ge 0$. In that case, as $\delta_3 \leq 1$, $\delta_3 \left[ z_i - \frac{Z}{n} \right] \le z_i - \frac{Z}{n} \leq z_i$, as desired.

    As for \emph{need monotonicity}, let $(y,z),(y,z') \in \mathcal{D}^{N}$ and let $i \in N$ be such that $z_i>z'_i$. Then,
    $$
    R_i(y,z)-R_i(y,z') = (1-\delta_3)\left[(z_i-z'_i)-\frac{Z-Z'}{n} \right] = (1-\delta_3)(z_i-z'_i) \frac{n-1}{n}.
    $$
    As $\delta_3 \le 1$ and $z_i > z'_i$, we conclude that $R_i(y,z)-R_i(y,z') \ge 0$, as desired.
\end{itemize}
We now focus on the converse implications. First, let $R$ be a rule that satisfies \textit{equal treatment of equals}, \textit{additivity}, and either \emph{needs lower bound} or \emph{net-average lower bound}. Let $(y,z) \in \mathcal{D}^N$. By \emph{additivity},
$$
R(y,z) = R(y-z,0) + R(z,z).
$$
Let $i \in N$. The combination of \emph{additivity} and one of the boundedness axioms implies that (e.g. Acz\'{e}l and Dhombres, 1989, Ch. 4)
$$
R_i(y-z,0)=\sum_{k=1}^{n}  R_i((y_{k}-z_{k},0_{-k}),0)=\sum_{k=1}^{n}  (y_{k}-z_{k}) R_i((1_k,0_{-k}),0).
$$
and 
$$
R_i(z,z) =\sum_{k=1}^{n}  z_{k} R_i((1_k,0_{-k}),(1_k,0_{-k})).
$$
We can now reproduce the same argument as in the proof of Theorem \ref{ADD_AN} to obtain that 
$$
R(y,z) = \lambda_1 R^{L}(y,z) + \lambda_2 R^{F}(y,z) + (1 - \lambda_1 - \lambda_2) R^{A}(y,z),
$$
where $\lambda_1=\frac{n\alpha-1}{n-1}$, $\lambda_2=\frac{n(1-\beta)}{n-1}$, and, therefore, $1 - \lambda_1 - \lambda_2=\frac{n(\beta-\alpha)}{n-1}$. In addition, $\alpha=R_k((1_k,0_{-k}),0)$ and $\beta=R_k((1_k,0_{-k}),(1_k,0_{-k}))$ for each $k \in N$. We now distinguish three cases.
\begin{enumerate}
  \item[(a)] Suppose that $R$ also satisfies \textit{zero lower bound}, \emph{order preservation for uniform needs}, and \emph{strong need monotonicity}. 
  
  By \emph{strong need monotonicity}, $R_k((1_k,0_{-k}),(1_k,0_{-k})) \geq R_k((1_k,0_{-k}),0)$. Thus, $\beta \geq \alpha$. 

  By \emph{strong need monotonicity}, for each $j \in N \backslash \{k\}$, $R_j((1_k,0_{-k}),(1_k,0_{-k})) \geq R_j((1_k,0_{-k}),0)$. By this, together with \emph{equal treatment of equals}, $\frac{1-\beta}{n-1} \geq \frac{1-\alpha}{n-1}$, or equivalently, $\beta \leq \alpha$. 
  
  Therefore, we conclude that $\alpha=\beta$, and then $1 - \lambda_1 - \lambda_2=0$.

  By \emph{zero lower bound}, for each $j \in N\backslash \{k\}$, $R_j((1_k,0_{-k}),0) \geq 0$. As $\sum_{k=1}^n R_k((1_k,0_{-k}),0)=1$, we conclude that $\alpha=R_k((1_k,0_{-k}),0) \leq 1$. 
  
  Finally, by \emph{order preservation for uniform needs} and \emph{equal treatment of equals}, for each $j \in N \backslash \{k\}$ 
  $R_k((1_k,0_{-k}),0) \geq R_j((1_k,0_{-k}),0)$. Or, equivalently, $\alpha \geq \frac{1-\alpha}{n-1}$. That is, $\alpha \geq \frac{1}{n}$. 
  
  Thus, if we let $\delta_1=\lambda_1=\frac{n\alpha-1}{n-1}$, it follows that $\delta_1 \in [0,1]$, and
  $$
  R(y,z) = \delta_1 R^{L}(y,z) + (1-\delta_1) R^{F}(y,z),
  $$
  as desired. 

  \item[(b)] Suppose instead that $R$ also satisfies \textit{needs lower bound} and \emph{order preservation for uniform needs}. 
  
  By \emph{needs lower bound}, $\beta=R_k((1_k,0_{-k}),(1_k,0_{-k})) \geq 1$ and, for each $j \in N\setminus\{k\}$, $R_j((1_k,0_{-k}),(1_k,0_{-k})) \geq 0$. As $\sum_{j=1}^n R_j(1_k,0_{-k}),(1_k,0_{-k})=1$, we conclude that $\beta=R_k((1_k,0_{-k}),(1_k,0_{-k})) \leq 1$. Thus, $\beta=1$ and, therefore, $\lambda_2=0$.

  By \emph{needs lower bound}, for each $j \in N\setminus\{k\}$, $R_j((1_k,0_{-k}),0) \geq 0$. As $\sum_{j=1}^n R_j((1_k,0_{-k}),0)=1$, we conclude that $\alpha=R_k((1_k,0_{-k}),0) \leq 1$. 
  
  On the other hand, by \emph{equal treatment of equals}, for each $j \in N\setminus\{k\}$, $R_j((1_k,0_{-k}),0)=\frac{1-\alpha}{n-1}$. And, by \emph{order preservation for uniform needs}, $\alpha \ge \frac{1-\alpha}{n-1}$ or, equivalently, $\alpha \ge \frac{1}{n}$. 
  
  Thus, if we let $\delta_2=\lambda_1=\frac{n\alpha-1}{n-1}$, it follows that $\delta_2 \in [0,1]$, and
  $$
  R(y,z) = \delta_2 R^{L}(y,z) + (1-\delta_2) R^{A}(y,z),
  $$
  as desired.

  \item[(c)] Finally, suppose instead that $R$ also satisfies \textit{net-average lower bound} and \emph{need monotonicity}.
  
  By \emph{net-average lower bound}, for each $j \in N\setminus\{k\}$, $R_j((1_k,0_{-k}),0) \ge \frac{1}{n} \geq 0$. As $\sum_{j=1}^n R_j((1_k,0_{-k}),0) = 1$, we conclude that $\alpha=R_k((1_k,0_{-k}),0)=\frac{1}{n}$ and, therefore, $\lambda_1=0$.

  By \emph{net-average lower bound}, we also obtain that, for each $j \in N\setminus\{k\}$, $R_j((1_k,0_{-k}),(1_k,0_{-k})) \ge \frac{1}{n}$. As $\sum_{j=1}^n R_j((1_k,0_{-k}),(1_k,0_{-k})) = 1$, we conclude that $\beta=R_k((1_k,0_{-k}),(1_k,0_{-k})) \leq 1$. 
  
  Finally, by \emph{need monotonicity}, it follows that $\beta \ge \frac{1}{n}$. 
  
  Thus, if we let $\delta_3=\lambda_2=\frac{n(1-\beta)}{n-1}$, it follows that $\delta_3 \in [0,1]$, and
  $$
  R(y,z) = \delta_3 R^{F}(y,z) + (1-\delta_3) R^{A}(y,z),
  $$
  as desired. \qed
\end{enumerate}

\subsection*{Proof of Theorem \ref{char_focal}}

We already know from Theorem \ref{ADD_AN_THIRD} that each rule satisfies the axioms in the corresponding statement. We focus on the converse implication. 
\begin{enumerate}
  \item[(a)] Suppose that $R$ satisfies \emph{additivity}, \emph{needs lower bound} and \emph{strong need monotonicity}. Let $(y,z) \in \mathcal{D}^N$. By \emph{additivity},
  $$
  R(y,z) = R(y-z,0) + R(z,z).
  $$
  Let $i \in N$. By \emph{additivity} and \emph{needs lower bound} (e.g., Acz\'{e}l and Dhombres, 1989, Ch. 4),
  $$
  R_i(y-z,0)=\sum_{k=1}^{n}  R_i((y_{k}-z_{k},0_{-k}),0)=\sum_{k=1}^{n}  (y_{k}-z_{k}) R_i((1_k,0_{-k}),0),
  $$
  and 
  $$
  R_i(z,z) =\sum_{k=1}^{n}  z_{k} R_i((1_k,0_{-k}),(1_k,0_{-k})).
  $$
  By \emph{needs lower bound}, 
  $R((1_k,0_{-k}),(1_k,0_{-k})) \geq (1_k,0_{-k}) \geq 0$. Thus, by the definition of rules, $R((1_k,0_{-k}),(1_k,0_{-k})) = (1_k,0_{-k})$. 
  Now, by \emph{strong need monotonicity}, $R((1_k,0_{-k}),(1_k,0_{-k})) \geq R((1_k,0_{-k}),0)$, or, equivalently, $(1_k,0_{-k}) \geq R((1_k,0_{-k}),0)$. In addition, by  \emph{needs lower bound}, $R((1_k,0_{-k}),0) \geq 0$. As $(1_k,0_{-k}) \geq R((1_k,0_{-k}),0) \geq (0_k,0_{-k})$, it follows that $R((1_k,0_{-k}),0)=(1_k,0_{-k})$. Therefore,
  \begin{align*}
    R_i(y,z) &= \sum_{k=1}^{n}  (y_{k}-z_{k}) R_i((1_k,0_{-k}),0) +  \sum_{k=1}^{n}  z_{k} R_i((1_k,0_{-k}),(1_k,0_{-k})) \\
             &= (y_i - z_i) + z_i \\
             &= y_i.
  \end{align*}
  \item[(b)] Suppose that $R$ satisfies \emph{additivity}, \emph{net-average lower bound} and \emph{strong need monotonicity}. Let $(y,z) \in \mathcal{D}^N$. By \emph{additivity} and \emph{net-average lower bound}, replicating the argument in (a), we have 
  $$
  R_i(y,z) = \sum_{k=1}^{n}  (y_{k}-z_{k}) R_i((1_k,0_{-k}),0) +  \sum_{k=1}^{n}  z_{k} R_i((1_k,0_{-k}),(1_k,0_{-k})).
  $$
  By \emph{net-average lower bound}, 
  $R((1_k,0_{-k}),0) \geq \left(\frac{1}{n},\ldots,\frac{1}{n} \right)$. Thus, by the definition of rules, 
  $R((1_k,0_{-k}),0) = \left(\frac{1}{n},\ldots,\frac{1}{n} \right)$. Now, by \emph{strong need monotonicity}, $R((1_k,0_{-k}),(1_k,0_{-k})) \geq R((1_k,0_{-k}),0) = \left(\frac{1}{n},\ldots,\frac{1}{n} \right)$. Thus, by the definition of rules, $R((1_k,0_{-k}),(1_k,0_{-k})) = \left(\frac{1}{n},\ldots,\frac{1}{n} \right)$. Therefore,
  \begin{align*}
    R_i(y,z) &= \sum_{k=1}^{n}  (y_{k}-z_{k}) R_i((1_k,0_{-k}),0) +  \sum_{k=1}^{n}  z_{k} R_i((1_k,0_{-k}),(1_k,0_{-k})) \\
             &= \sum_{k=1}^{n}  (y_{k}-z_{k}) \frac{1}{n} +  \sum_{k=1}^{n}  z_{k} \frac{1}{n} \\
             &= \frac{Y}{n}.
  \end{align*}
  \item[(c)] Suppose that $R$ satisfies \emph{additivity}, \emph{needs monotonicity} and \emph{net-average lower bound}. Let $(y,z) \in \mathcal{D}^N$. By \emph{additivity} and \emph{needs monotonicity} (replicating the arguments above) we have that 
  $$
  R_i(y,z) = \sum_{k=1}^{n}  (y_{k}-z_{k}) R_i((1_k,0_{-k}),0) +  \sum_{k=1}^{n}  z_{k} R_i((1_k,0_{-k}),(1_k,0_{-k})).
  $$
  As in (a), \emph{needs monotonicity} implies that $R((1_k,0_{-k}),(1_k,0_{-k})) = (1_k,0_{-k})$. 
  And, as in (b), \emph{net-average lower bound} implies that $R((1_k,0_{-k}),0) = \left(\frac{1}{n},\ldots,\frac{1}{n} \right)$. Therefore,
  \begin{align*}
    R_i(y,z) &= \sum_{k=1}^{n}  (y_{k}-z_{k}) R_i((1_k,0_{-k}),0) +  \sum_{k=1}^{n}  z_{k} R_i((1_k,0_{-k}),(1_k,0_{-k})) \\
             &= \sum_{k=1}^{n}  (y_{k}-z_{k}) \frac{1}{n} + z_i \\
             &= z_i + \frac{Y-Z}{n}.
  \end{align*}
  \qed
\end{enumerate}

    \subsection*{Proof of Proposition \ref{Majority 1}}
        Let $\delta_1 \in [0,1]$, and $(y,z)\in \mathcal{D}^N$. For each $i\in N$, 
        \begin{equation*}
        R_{i}^{\delta_1}(y,z)=\delta_1 y_i + (1-\delta_1) \frac{Y}{n}=\delta_1(y_i-\frac{Y}{n})+\frac{Y}{n} .
        \end{equation*}
        
        If $y_{i}>\frac{Y}{n}$, then $R_{i}^{\delta_1}(y,z)$ is an increasing
        function of $\delta_1 $, thus maximized at $\delta_1 =1$.
        This implies that, for each $i\in N^{y}_{u}$, $R_{i}^{1}(y,z)= R^{L}(y,z)$ is the most preferred outcome.
        
        If $y _{i}<\frac{Y}{n}$, then $R_{i}^{\delta_1}(y,z)$ is a decreasing
        function of $\delta_1 $, thus maximized at $\delta_1 =0$.
        This implies that, for each $i\in N^{y}_{l}$, $R_{i}^{0}(y,z)= R^{F}(y,z)$ is the most preferred outcome.
        
        If $y _{i}=\frac{Y}{n},$ then $R_{i}^{\delta_1}(y,z)=\frac{Y}{n}$ for each $\delta_1 \in [0,1] $. This implies that, for each $i\in N^{y}_{e}$,
        all rules within the family $\left\{ R^{\delta_1}\right\} _{\delta_1 \in [0,1]}$ yield the same outcome.
        
        From the above, statements $\left( i\right) $ and $\left( ii\right) $ follow trivially. Assume, by contradiction, that statement $\left( iii\right) $ does not hold. Then, there exists $(y,z)\in \mathcal{D}^N$ and $\delta_1 \in [0,1] $ such that $R^{\delta_1}(y,z)$ is not a majority winner.
        Thus, we can find $\delta_1 ^{\prime }\in [0,1]$ such that $R_{i}^{\delta_1 ^{\prime }}\left( y,z\right)
        >R^{\delta_1}_{i}\left( y,z\right) $ holds for the majority of the agents. We
        then consider two cases: 
        
        \textbf{Case $\delta_1 ^{\prime }>\delta_1 $}.
        
        In this case, $R_{i}^{\delta_1  ^{\prime }}(y,z) >R_{i}^{\delta_1 
        }(y,z) $ if and only if $i\in N^{y}_{u}.$ Now, 
        $$
        \left\vert N^{y}_{u}\right\vert =\left\vert \left\{ i\in N:R_{i}^{\delta_1 
        ^{\prime }}(y,z) >R_{i}^{\delta_1  }(y,z) \right\}
        \right\vert  > \left\vert \left\{ i\in N:R_{i}^{\delta_1  ^{\prime }}(y,z) \leq
        R_{i}^{\delta_1  }(y,z) \right\} \right\vert = |N^{y}_{l}|+\left\vert N^{y}_{e}\right\vert,
        $$
        which is a contradiction.%
        
        \textbf{Case $\delta_1  ^{\prime }<\delta_1  $}.
        
        In this case, $R_{i}^{\delta_1  ^{\prime }}(y,z) >R_{i}^{\delta_1 
        }(y,z) $ if and only if $i\in N^{y}_{l}.$ Now, 
        $$
        \left\vert N^{y}_{l}\right\vert =\left\vert \left\{ i\in N:R_{i}^{\delta_1 
        ^{\prime }}(y,z) >R_{i}^{\delta_1  }(y,z) \right\}
        \right\vert 
        >\left\vert \left\{ i\in N:R_{i}^{\delta_1  ^{\prime }}(y,z) \leq
        R_{i}^{\delta_1  }(y,z) \right\} \right\vert = |N^{y}_{u}|+\left\vert N^{y}_{e}\right\vert,
        $$
        which is a contradiction. \qed
        \subsection*{Proof of Proposition \ref{Majority 2}}
            Let $\delta_2 \in [0,1]$, and $(y,z)\in \mathcal{D}^N$. For each $i\in N$, 
            $$R^{\delta_2}_i(y,z)=\delta_2 y_i + (1-\delta_2)\left[z_i + \frac{Y-Z}{n} \right]=\delta_2 \left[y_i -z_i - \frac{Y-Z}{n} \right]+\left[z_i + \frac{Y-Z}{n} \right].$$
            
            If $y_{i}-z_{i}>\frac{Y-Z}{n}$, then $R_{i}^{\delta_2}(y,z)$ is an increasing
            function of $\delta_2 $, thus maximized at $\delta_2 =1$.
            This implies that, for each $i\in N_{u}$, $R_{i}^{1}(y,z)= R^{L}(y,z)$ is the most preferred outcome.
            
            If $y_{i}-z_{i}<\frac{Y-Z}{n}$, then $R_{i}^{\delta_2}(y,z)$ is a decreasing
            function of $\delta_2 $, thus maximized at $\delta_2 =0$.
            This implies that, for each $i\in N_{l}$, $R_{i}^{0}(y,z)= R^{A}(y,z)$ is the most preferred outcome.
            
            If $y_{i}-z_{i}=\frac{Y-Z}{n},$ then $R_{i}^{\delta_2}(y,z)=z_i + \frac{Y-Z}{n}$ for each $\delta_2 \in [0,1] $. This implies that, for each $i\in N_{e}$,
            all rules within the family $\left\{ R^{\delta_2}\right\} _{\delta_2 \in [0,1]}$ yield the same outcome.
            
            From the above, statements $\left( i\right) $ and $\left( ii\right) $ follow trivially. Assume, by contradiction, that statement $\left( iii\right) $ does not hold. Then, there exists $(y,z)\in \mathcal{D}^N$ and $\delta_2 \in [0,1] $ such that $R^{\delta_2}(y,z)$ is not a majority winner.
            Thus, we can find $\delta_2 ^{\prime }\in [0,1]$ such that $R_{i}^{\delta_2 ^{\prime }}\left( y,z\right)
            >R^{\delta_2}_{i}\left( y,z\right) $ holds for the majority of the agents. We
            then consider two cases: 
            
            \textbf{Case $\delta_2 ^{\prime }>\delta_2 $}.
            
            In this case, $R_{i}^{\delta_2  ^{\prime }}(y,z) >R_{i}^{\delta_2 
            }(y,z) $ if and only if $i\in N_{u}.$ Now, 
            $$
            \left\vert N_{u}\right\vert =\left\vert \left\{ i\in N:R_{i}^{\delta_2 
            ^{\prime }}(y,z) >R_{i}^{\delta_2  }(y,z) \right\}
            \right\vert 
            > \left\vert \left\{ i\in N:R_{i}^{\delta_2  ^{\prime }}(y,z) \leq
            R_{i}^{\delta_2  }(y,z) \right\} \right\vert 
            =|N_{l}|+\left\vert N_{e}\right\vert,
            $$
            which is a contradiction.%
            
            \textbf{Case $\delta_2  ^{\prime }<\delta_2  $}.
            
            In this case, $R_{i}^{\delta_2  ^{\prime }}(y,z) >R_{i}^{\delta_2 
            }(y,z) $ if and only if $i\in N_{l}.$ Now, 
            $$
            \left\vert N_{l}\right\vert =\left\vert \left\{ i\in N:R_{i}^{\delta_2 
            ^{\prime }}(y,z) >R_{i}^{\delta_2  }(y,z) \right\}
            \right\vert  > \left\vert \left\{ i\in N:R_{i}^{\delta_2  ^{\prime }}(y,z) \leq
            R_{i}^{\delta_2  }(y,z) \right\} \right\vert = |N_{u}|+\left\vert N_{e}\right\vert,
            $$
            which is a contradiction. \qed
\subsection*{Proof of Proposition \ref{Majority 3}}
                Let $\delta_3 \in [0,1]$, and $(y,z)\in \mathcal{D}^N$. For each $i\in N$, 
                $$R^{\delta_3}_i(y,z)=
                \delta_3 \frac{Y}{n} + (1-\delta_3) \left[ z_i + \frac{Y-Z}{n}\right]=z_i + \frac{Y-Z}{n}-\delta_3 \left[ z_i - \frac{Z}{n}\right].$$
                
                If $z_{i}<\frac{Z}{n}$, then $R_{i}^{\delta_3}(y,z)$ is an increasing
                function of $\delta_3 $, thus maximized at $\delta_3 =1$.
                This implies that, for each $i\in N^{z}_{l}$, $R_{i}^{1}(y,z)= R^{F}(y,z)$ is the most preferred outcome.
                
                If $z_{i}>\frac{Z}{n}$, then $R_{i}^{\delta_3}(y,z)$ is a decreasing
                function of $\delta_3 $, thus maximized at $\delta_3 =0$.
                This implies that, for each $i\in N^{z}_{u}$, $R_{i}^{0}(y,z)= R^{A}(y,z)$ is the most preferred outcome.
                
                If $z_{i}=\frac{Z}{n},$ then $R_{i}^{\delta_3}(y,z)=z_i + \frac{Y-Z}{n}$ for each $\delta_3 \in [0,1] $. This implies that, for each $i\in N^{z}_{e}$,
                all rules within the family $\left\{ R^{\delta_3}\right\} _{\delta_3 \in [0,1]}$ yield the same outcome.
                
                From the above, statements $\left( i\right) $ and $\left( ii\right) $ follow trivially. Assume, by contradiction, that statement $\left( iii\right) $ does not hold. Then, there exists $(y,z)\in \mathcal{D}^N$ and $\delta_3 \in [0,1] $ such that $R^{\delta_3}(y,z)$ is not a majority winner.
                Thus, we can find $\delta_3 ^{\prime }\in [0,1]$ such that $R_{i}^{\delta_3 ^{\prime }}\left( y,z\right)
                >R^{\delta_3}_{i}\left( y,z\right) $ holds for the majority of the agents. We
                then consider two cases: 
                
                \textbf{Case $\delta_3 ^{\prime }>\delta_3 $}.
                
                In this case, $R_{i}^{\delta_3  ^{\prime }}(y,z) >R_{i}^{\delta_3 
                }(y,z) $ if and only if $i\in N^{z}_{l}.$ Now, 
                $$\left\vert N^{z}_{l}\right\vert 
                =\left\vert \left\{ i\in N:R_{i}^{\delta_3 
                ^{\prime }}(y,z) >R_{i}^{\delta_3  }(y,z) \right\}
                \right\vert 
                >\left\vert \left\{ i\in N:R_{i}^{\delta_3  ^{\prime }}(y,z) \leq
                R_{i}^{\delta_2  }(y,z) \right\} \right\vert 
                =|N^{z}_{u}|+\left\vert N^{z}_{e}\right\vert,
                $$
                which is a contradiction.%
                
                \textbf{Case $\delta_3  ^{\prime }<\delta_3  $}.
                
                In this case, $R_{i}^{\delta_3  ^{\prime }}(y,z) >R_{i}^{\delta_3 
                }(y,z) $ if and only if $i\in N^{z}_{u}.$ Now, 
                $$
                \left\vert N^{z}_{u}\right\vert =\left\vert \left\{ i\in N:R_{i}^{\delta_3 
                ^{\prime }}(y,z) >R_{i}^{\delta_3  }(y,z) \right\}
                \right\vert >
                \left\vert \left\{ i\in N:R_{i}^{\delta_3  ^{\prime }}(y,z) \leq
                R_{i}^{\delta_3  }(y,z) \right\} \right\vert = |N^{z}_{l}|+\left\vert N^{z}_{e}\right\vert,
                $$
                which is a contradiction. \qed
                \subsection*{Proof of Proposition \ref{Lorenz 1}}
                    \begin{itemize}
                    \item[(a)] Let $(y,z) \in \mathcal{D}^N$. Let us assume, without loss of generality, that $y_1 \le \ldots \le y_n$. 
                    Then, $R^\delta_1(y,z) \le \ldots \le R^\delta_n(y,z)$, for each $\delta \in [0,1]$. 
                    Let $\delta, \overline{\delta} \in [0,1]$ be such that $\delta \le \overline{\delta}$. Then, for each $k \in \{1,\ldots,n\}$,
                    \begin{align*}
                         S^k \left( R^\delta(y,z) \right) \ge S^k \left( R^{\overline{\delta}}(y,z) \right) 
                         &\Leftrightarrow \sum_{i=1}^k \left[ \delta \left[ y_i - \frac{Y}{n} \right] + \frac{Y}{n} \right] \ge \sum_{i=1}^k \left[ \overline{\delta} \left[ y_i - \frac{Y}{n} \right] + \frac{Y}{n} \right] \\
                         &\Leftrightarrow \delta \left[ \sum_{i=1}^k y_i -k \frac{Y}{n} \right] \ge \overline{\delta} \left[ \sum_{i=1}^k y_i -k \frac{Y}{n} \right] \\
                         &\Leftrightarrow (\delta-\overline{\delta}) \left[ \sum_{i=1}^k y_i -k \frac{Y}{n} \right] \ge 0
                    \end{align*}
                    As $\delta \leq \overline{\delta}$ and $kY \ge n \sum_{i=1}^k y_i$, the previous condition holds. Thus, $R^\delta(y,z) \succeq_{L} R^{\overline{\delta}}(y,z)$, as desired. 
                    \item[(b)] Let $(y,z) \in \mathcal{D}^N$. Let us assume, without loss of generality, that $z_1 \le \ldots \le z_n$. 
                    Then, $R^\delta_1(y,z) \le \ldots \le R^\delta_n(y,z)$. 
                      Let $\delta, \overline{\delta} \in [0,1]$ be such that $\delta \le \overline{\delta}$. Then, for each $k \in \{1,\ldots,n\}$,
                      \begin{align*}
                        S^k\left( R^{\overline{\delta}}(y,z) \right) \ge S^k \left( R^\delta(y,z) \right)
                        &\Leftrightarrow \sum_{i=1}^k \left[ \overline{\delta} \left[ \frac{Z}{n} - z_i \right] + z_i + \frac{Y-Z}{n} \right] \ge \sum_{i=1}^k \left[ \delta \left[ \frac{Z}{n} - z_i \right] + z_i + \frac{Y-Z}{n} \right] \\
                        &\Leftrightarrow \overline{\delta} \left[ k\frac{Z}{n} - \sum_{i=1}^k z_i \right] \ge \delta \left[ k\frac{Z}{n} - \sum_{i=1}^k z_i \right] \\
                        &\Leftrightarrow (\overline{\delta}-\delta) \left[ k\frac{Z}{n} - \sum_{i=1}^k z_i \right] \ge 0.
                      \end{align*}
                      As $\delta \le \overline{\delta}$ and $kZ \ge n \sum_{i=1}^k z_i$, the previous condition holds. Thus, $R^{\overline{\delta}}(y,z) \succeq_{L} R^\delta(y,z) $, as desired. \qed
                    \end{itemize}
                    \subsection*{Proof of Proposition \ref{Lorenz 2}}
                        Consider the problem $(y,z) \in \mathcal{D}^N$ where $y=(2,2,10)$ and $z=(1,4,0)$. Then, by definition,
                        $$
                        R^\frac{1}{3}(y,z)=\frac{1}{3}(2,2,10)+\frac{2}{3}(4, 7,3)=\left(\frac{10}{3}, \frac{16}{3}, \frac{16}{3}\right),
                        $$
                        and
                        $$
                        R^\frac{1}{10}(y,z)=\frac{1}{10}(2,2,10)+\frac{9}{10}(4,7,3)=\left(\frac{38}{10}, \frac{65}{10}, \frac{37}{10}\right).
                        $$
                        Notice that $S^1(R^\frac{1}{3}(y,z))=\frac{10}{3} < \frac{37}{10} = S^1(R^\frac{1}{10}(y,z))$ and $S^2(R^\frac{1}{3}(y,z))=\frac{26}{3} > \frac{75}{10} = S^2(R^\frac{1}{10}(y,z))$. Therefore, neither $R^\frac{1}{3}(y,z) \succeq_L R^\frac{1}{10}(y,z)$ nor $R^\frac{1}{10}(y,z) \succeq_L R^\frac{1}{3}(y,z)$. \qed


\begin{thebibliography}{99}

\bibitem{} Acz\'{e}l, J., (2006) Lectures on functional equations and their applications. Dover.

\bibitem{} Acz\'{e}l, J., Dhombres, J. (1989). Functional Equations in Several Variables. Cambridge University Press.

\bibitem{} Alm\r{a}s, I., Cappelen, A., Thori Lind, J., S\o rensen, E., Tungodden, B., (2011) Measuring unfair (in) equality. Journal of Public Economics 95, 488-499.

\bibitem{} Atkinson, A., Bourguignon, F., (1987) Income Distributions and Differences in Needs, in Arrow and the Foundation of the Theory of Economic Policy, ed. by G. R. Feiwel. Macmillan, London.

\bibitem{} Atkinson, A., Stiglitz, J., (1980) Lectures on Public Economics. McGraw-Hill, New York.


\bibitem{BMT20a} Berganti\~{n}os, G., Moreno-Ternero, J.D., (2020) Sharing the revenues from broadcasting sport events. Management Science 66, 2417-2431.

\bibitem{BMT21a} Berganti\~{n}os, G., Moreno-Ternero, J.D., (2021) Compromising to share the revenues from broadcasting sports leagues. Journal of Economic Behavior and Organization 183, 57-74.

\bibitem{BMT22c} Berganti\~{n}os, G., Moreno-Ternero, J.D., (2022) On the axiomatic approach to sharing the revenues from broadcasting sports leagues. Social Choice and Welfare 58, 321-347.


\bibitem{BMT22d} Berganti\~{n}os, G., Moreno-Ternero, J.D., (2023) Decentralized revenue sharing from broadcasting sports. Public Choice 194, 27-44.

\bibitem{} Berganti\~{n}os, G., Vidal-Puga, J., (2004) Additive rules in bankruptcy problems and other related problems. Mathematical Social Sciences 47, 87-101.


\bibitem{} Billette de Villemeur, E., Leroux, J., (2023). Accounting for needs when sharing costs. Mimeo.


\bibitem{} Bourguignon, F., (1989) Family Size and Social Utility: Income Distribution Dominance Criteria. Journal of Econometrics 42, 67-80.

\bibitem{} Calabrese, S., (2007) Majority voting over publicly provided goods, redistribution, and income taxation. Journal of Public Economic Theory 9, 319-334.

\bibitem{} Casajus, A., (2015) Monotonic redistribution of performance-based allocations: A case for proportional taxation. Theoretical Economics 10, 887-892.



\bibitem{} Chambers, C.P., Moreno-Ternero, J.D., (2017) Taxation and Poverty. Social Choice and Welfare 48, 153-175.

\bibitem{} Chambers, C.P., Moreno-Ternero, J.D., (2021) Bilateral redistribution. Journal of Mathematical Economics 96, 102517.

\bibitem{} Chun, Y., (1988) The proportional solution for rights problems. Mathematical Social Sciences 15, 231-246.

\bibitem{} de Clippel, G., Rozen, K., (2022) Fairness through the lens of cooperative game theory: an experimental approach. American Economic Journal: Microeconomics 14, 810-36.

\bibitem{} Corch\'{o}n L., Puy, S., (1998) Individual rationality and voting in cooperative production. Economics Letters 59, 83-90.


\bibitem{} Dasgupta, P.S., Sen, A.K., Starret, D., (1973) Notes on the measurement of inequality. Journal of Economic Theory 6, 180-187.

\bibitem{} Dixit, A., Sandmo, A., (1977) Some simplified formulae for optimal income taxation. Scandinavian Journal of Economics 79, 417-423.


\bibitem{} Eichorn, W., (1978) Functional Equations in Economics. Addison-Wesley, Reading, Massachusetts.

\bibitem{} Faure, M., Gravel, N., (2021) Reducing Inequalities among Unequals. International Economic Review 62, 357-404.

\bibitem{} Gans, J., Smart, M., (1996) Majority voting with single-crossing preferences. Journal of Public Economics 59, 219-237.


\bibitem{} Greenberg, J., (1979) Consistent majority rules over compact sets of alternatives. Econometrica 47, 627-636.

\bibitem{} Hamada, K., (1973) A simple majority rule on the distribution of income. Journal of Economic Theory 6, 243-264.

\bibitem{} Hemming, R., Keen, M., (1983) Single-crossing conditions in comparisons of tax progressivity. Journal of Public Economics 20, 373-380.


\bibitem{} Hougaard, J.L., Moreno-Ternero, J.D., \O sterdal, L.P., (2012) A unifying framework for the problem of adjudicating conflicting claims. Journal of Mathematical Economics 48, 107-114.

\bibitem{} Hougaard, J., Moreno-Ternero, J.D., \O sterdal, L.P., (2013) A new axiomatic approach to the evaluation of population health. Journal of Health Economics 32, 515-523.
    


\bibitem{} Hougaard, J.L., Moreno-Ternero, J.D., Tvede, M., \O sterdal, L.P., (2017) Sharing the proceeds from a hierarchical venture. Games and Economic Behavior 102, 98-110.

\bibitem{} Ihori, T., (1987) The optimal linear income tax: a diagrammatic analysis, Journal of Public Economics 34, 379-390.

\bibitem{} Jakobsson, U., (1976) On the measurement of the degree of progression. Journal of Public Economics 5, 161-168.

\bibitem{Ju et al} Ju, B.-G., Miyagawa E., Sakai T., (2007) Non-manipulable division rules in claim problems and generalizations. Journal of Economic Theory 132, 1-26.

\bibitem{} Ju, B-G., Moreno-Ternero, J.D., (2011) Progressive and merging-proof taxation. International Journal of Game Theory 40, 43-62.

\bibitem{} Ju, B-G., Moreno-Ternero, J.D., (2023) Taxation behind the veil of ignorance. Social Choice and Welfare 60, 165-181.




\bibitem{} Martinez, R., Moreno-Ternero J.D., (2022) Laissez-faire or full redistribution? Economics Letters 218, 110756.


\bibitem{} Martinez, R., Sanchez-Soriano J., (2023) Order preservation with dummies in the museum pass problem. Annals of Operations Research, forthcoming.


\bibitem{} Mirrlees, J.A., (1971) An exploration in the theory of optimal income taxation, Review of
Economic Studies 38, 175-208.

\bibitem{} Moreno-Ternero, J.D., Platz, T.T., \O sterdal, L.P., (2023) QALYs, DALYs, and HALYs: a unifying framework for the evaluation of population health. Journal of Health Economics 87, 102714.

\bibitem{} Moreno-Ternero J.D., Roemer J., (2006) Impartiality, priority, and solidarity and in the theory of justice. Econometrica 74, 1419-1427.


\bibitem{} Moreno-Ternero, J.D., \O sterdal, L.P., (2017) A normative foundation for equity-sensitive health evaluation: the role of relative comparisons of health gains. Journal of Public Economic Theory 19, 1009-1025.






\bibitem{} Moulin, H. (1987). Equal or proportional division of a surplus, and other methods. International Journal of Game Theory 16, 161-186.


\bibitem{} Nussbaum, M., Sen, A. (1993) The quality of life. Clarendon Press.


\bibitem{} O'Neill B., (1982) A problem of rights arbitration from the Talmud. Mathematical Social Sciences 2, 345-371.

\bibitem{} Roberts, K., (1977) Voting over income tax schedules. Journal of Public Economics 8, 329-340.

\bibitem{} Romer, T., (1975) Individual welfare, majority voting, and the properties of a linear income tax. Journal of Public Economics 4, 163-186.

\bibitem{} Sen, A., (1999) Commodities and capabilities. Oxford University Press.

\bibitem{} Shapley, L., (1953) A value for n-person games, in Contributions to the Theory of Games II (Annals of Mathematics Studies 28), ed. by H.W. Kuhn and A.W. Tucker, Princeton: Princeton University Press, 307-317.

\bibitem{} Stone, R., (1954), Linear expenditure systems and demand analysis: An application to the pattern of British demand. Economic Journal 64, 511-527.


\bibitem{key-9}Thomson, W., (2011), Fair allocation rules, Chapter
21 in the Handbook of Social Choice and Welfare Vol. 2, Arrow, K.,
Sen, A., Suzumura, K., eds. North Holland, 393-506 



\bibitem{} Thomson, W., (2019) How to divide when there isn't enough: from Aristotle, the Talmud, and Maimonides to the axiomatics of resource allocation. Econometric Society Monograph. Cambridge University Press. Cambridge, MA.

\bibitem{} Thomson, W., (2023) The Axiomatics of Economic Design, Vol. 1. An Introduction to Theory and Methods. Springer, Cham.

\bibitem{} Van Parijs, P., Vanderborght, Y., (2017) Basic income: a radical proposal for a free society and a sane economy. Harvard University Press. Cambridge, MA.



\bibitem{} Young H.P., (1988) Distributive justice in taxation. Journal of Economic Theory 44, 321-335.

\bibitem{} Young H.P., (1990) Progressive taxation and equal sacrifice. American Economic Review 80, 253-266.

\end{thebibliography}
\end{document}